\documentclass[journal]{IEEEtran}

\usepackage[pagewise, switch]{lineno} 

\ifCLASSINFOpdf
\else
\fi
\usepackage{xcolor}

\usepackage[nocompress]{cite}
\usepackage{url}
\usepackage{amsmath,amssymb,graphicx}
\usepackage{multirow}
\usepackage{array}

\newcommand{\thickhline}{%
	\noalign {\ifnum 0=`}\fi \hrule height 1pt
	\futurelet \reserved@a \@xhline
}
\newcolumntype{"}{@{\hskip\tabcolsep\vrule width 1pt\hskip\tabcolsep}}

\begin{document}
%
\title{Multi-Scale Deep Compressive Imaging}
%
%
%

\author{Thuong~Nguyen~Canh 
        and Byeungwoo~Jeon
\thanks{T. N. Canh was with the Department
of Electrical and Computer Engineering, Sungkyunkwan University, South Korea. He is now with the Institute for Datability Science, Osaka University, Japan. e-mail: ngcthuong@ids.osaka-u.ac.jp}.
\thanks{B. Jeon are with the Department of Electrical and Computer Engineering, Sungkyunkwan University, Korea.}
}

%
%

\markboth{Journal of \LaTeX\ Class Files,~Vol.~14, No.~8, August~2015}%
{Shell \MakeLowercase{\textit{et al.}}: Bare Demo of IEEEtran.cls for IEEE Journals}
%



\maketitle

\begin{abstract}
Recently, deep learning-based compressive imaging (DCI) has surpassed the conventional compressive imaging in reconstruction quality and faster running time. While multi-scale has shown superior performance over single-scale, research in DCI has been limited to single-scale sampling. 
Despite training with single-scale images, DCI tends to favor low-frequency components similar to the conventional multi-scale sampling, especially at low subrate. From this perspective, it would be easier for the network to learn multi-scale features with a multi-scale sampling architecture. In this work, we proposed a multi-scale deep compressive imaging (MS-DCI) framework which jointly learns to decompose, sample, and reconstruct images at multi-scale. A three-phase end-to-end training scheme was introduced with an initial and two enhance reconstruction phases to demonstrate the efficiency of multi-scale sampling and further improve the reconstruction performance. We analyzed the decomposition methods (including Pyramid, Wavelet, and Scale-space), sampling matrices, and measurements and showed the empirical benefit of MS-DCI which consistently outperforms both conventional and deep learning-based approaches. 
\end{abstract}


%
\IEEEpeerreviewmaketitle

\section{Introduction}

\IEEEPARstart{C}{ompressive} sensing (CS) is an emerging sampling technique to overcome the Nyquist-Shannon sampling theorem under the assumption of the signal sparsity \cite{Donoho06, Do12, LGan07, Duarte12, Yin10}. A CS measurement $y \in \mathbb{R}^m$, is captured from a sparse or compressible signal, $x \in \mathbb{R}^n,m \ll n$, via 
\begin{equation}
y = \Phi x+ \eta,
\end{equation}
where $\Phi \in \mathbb{R}^{m \times n}$ denotes the sampling matrix and $\eta$ represents the additive noise, and \textcolor{black}{the ratio $m/n$ is named subrate}. With a random matrix $\Phi$, CS enables a simple sampling, compressing encoder, and computational security \cite{Laska11, Canh19}. CS became an active research topic with numerous applications in medical imaging, image restoration, hyper-spectral imaging, wireless communication, etc. However, there are three major challenges in CS as follows. 

\textbf{Sampling complexity}. With a Gaussian matrix \cite{Donoho06}, signals are guaranteed to recover at a high probability but an extremely high burden on computation and storage. Tremendous effort has sought to alleviate the computational complexity with block-based \cite{LGan07}, separable \cite{Duarte12}, and structure sampling \cite{Do12, Yin10}. 
	
\textbf{Reconstruction quality}. Over the past decade, various reconstruction models have exploited image priors such as nonlocal \cite{Canh16}, low-rank approximation \cite{Zhang14}, etc. Signal priors are further utilized in the sampling side but limited to general prior (i.e., low-frequency prior) because of the non-availability of the to-be-sampled signal. Researchers developed multi-scale CS \cite{Tsaig06, Fowler11, Canh15} which captures more low-frequency components. However, sampling and reconstruction are often designed separately, thus limiting their performance.   
	
\textbf{Reconstruction complexity}. With linear projection sampling, CS shifts most of the complexity to the decoder, thereby demanding tremendous complexity for reconstruction \cite{Goldstein}. Also, conventional CS often utilizes image priors and iterative optimization thus making the reconstruction more time-consuming. Various efforts have sought to develop a fast method \cite{Dinh17} but failed to maintain high reconstruction quality.  

With the massive data and huge computation for training, deep learning (DL) has shown state-of-the-art performance in many image restoration tasks like image super resolution \cite{Timofte17, Liu19}, denoising \cite{Zhang17, Zhang19, Liu19, Want18}, etc. Modeling the sampling as a deep learning layer, deep learning-based compressive imaging (DCI) jointly learns to sense and reconstruct signals in an end-to-end framework \cite{Kul16} and improves the reconstruction quality and reduce the reconstruction complexity. However, DCI also faces the sampling complexity when dealing with high dimensional signals such as image/video. Therefore, research in DCI has been limited to single-scale, block-based CS \cite{Kul16, Yao17, Adler17, Shi17, Shi18, Shi19CVPR}. Recently, researchers start integrating image priors such as multi-scale (Wavelet decomposition \cite{Liu19, Kang17}), non-local structure \cite{Lefk17, Ahn17} in DL-based image restoration to further improve the reconstruction quality. However, most researches are focused on reconstruction and frequently overlook the importance of sampling. As a result, little attention has been paid to the sampling, especially multi-scale sampling.

\begin{figure}[!t]
	\centering
	\includegraphics[scale=0.43]{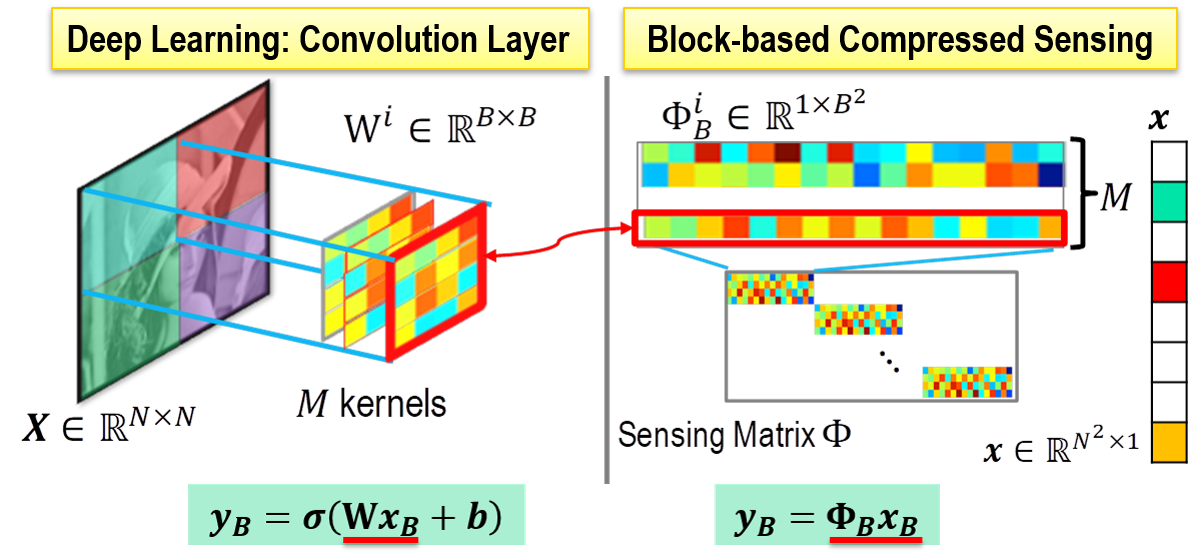}
	\caption{ \textcolor{black}{Comparison between BCS and convolution layer. Output of convolution layer is equivalent to BCS when $\Phi_B^i=vect(W^i)$, without bias $(b=0)$, and without activation $\sigma$. $\Phi_B$ is a block measurement matrix.}}
	\label{fig:dl_vs_bcs}
\end{figure}

\textcolor{black}{This work conduct the first analysis of the learned sampling matrices of single-scale DCI which results in aliasing artifact similar to the conventional multi-scale CS. Motivate by multi-scale CS \cite{Canh15, Canh18A}, we design a novel multi-scale architecture efficient sampling and reconstruct images at multi-scale.}   
 
In summary, we make the following contributions:
\begin{itemize}
	\item We propose an end-to-end multi-scale DCI with Pyramid, Wavelet, and Scale-space decomposition. We jointly learn to decompose, sample, and reconstruct images at multi-scale. 
	\item We present a three-phase training framework with initialization, and two enhance reconstruction networks.
	\item We analyze learned decomposition, learned sampling matrices, and demonstrate the high reconstruction quality of multi-scale DCI over state-of-the-art sensing schemes.
\end{itemize}
We also investigate the importance of non-linearity at the sampling stage. To reduce the sampling complexity, we use the combination of frame-based multi-scale decomposition, block-based sampling, and frame-based reconstruction. \textcolor{black}{This work focuses on deterministic sampling. Our multi-scale sampling matrices are fixed after the training process.} 


\section{Related Work}

\subsection{Deep Learning for Image Restoration }
Deep Learning (DL) for image restoration has been an active research area in recent years. Researchers started with a simple multiple layer perceptron \cite{Burger12}, then more advance architectures (i.e., CNN \cite{Zhang17}, residual learning \cite{Want18}, UNet-like networks \cite{Liu19}, \textcolor{black}{and generative model (GAN) \cite{Chen18,Ledig17}}). \textcolor{black}{The conventional restoration often utilizes internal (i.e., low-frequency, perceptual prior, non-local structure) more often than the external information. The dictionary learning approach does benefit from external information but on a small scale. Meanwhile, DL takes the advantage of a large scale dataset to learn the mapping from the degraded to clean image, also known as deep prior \cite{Chang17, Zhang19, Dmitry18}.} Despite outperforming the state-of-the-art conventional algorithms (sparse coding, low-rank approximation \cite{Canh16, Zhang14}), DL \cite{Liu19, Dmitry18, Zhang17} (excluding GAN) favor the external information more, thus loss quality for strong local structure images. GAN-based methods \cite{Ledig17}, on the other hand, created artificial-like structures in the reconstructed images which are not existing the original images. Thus, researchers integrated well-known priors to DL framework by \textcolor{black}{exploiting the iterative solution \cite{Dmitry18}}, nonlocal \cite{Lefk17, Ahn17}, low frequency prior \cite{Liu19, Kang17}, etc. 

\subsection{Compressive Sensing Meets Deep Learning}
\subsubsection{Block-based vs. Structure Frame-based}
To realize deep learning-based compressive imaging (DCI), the linear CS acquisition was modeled as a layer in the deep network. For instance, frame-based sensing is equivalent to a fully connected layer (FCL) without the activation and bias term \cite{Kul16}, block-based sampling as a convolution \cite{Shi17} with a large stride as in Fig. \ref{fig:dl_vs_bcs}. Since FCL-based DCI faces the sampling complexity issues as frame-based sensing, it was trained with small datasets of size $32\times 32$ or a small image block \cite{Kul16, Adler17}, thereby introduced the blocking artifact. Block-based DCI with frame-based recovery was preferred \cite{Shi17, Shi18} to reduce complexity without the blocking artifact. Also, DCI can learn to reconstruct images from measurements of fixed \cite{Kul16,Yao17, Ahn17} or learned sampling matrices \cite{Adler17, Shi17, Canh18A}. 
To reduce frame-based sensing complexity, Iliasdis et al. \cite{Iliadis16} learned a binary sampling matrix, Nguyen et al. \cite{Nguyen17} developed a sparse ternary matrix network with only values of {0, -1, +1}, Cui et al. \cite{Cui18} sparsified the learned sampling matrix and our previous work on separable DCI \cite{Canh18A}. 

\subsubsection{Single-Scale vs. Multiple-Scale }
Acquiring images in the original or multi-level decomposition domain (i.e. wavelet transform domain) are named single-scale or multi-scale sampling, respectively. While the advantage of multi-scale over signal scale has been well studied in the literature \cite{Tsaig06, Fowler11, Canh15, Canh18B}, most DCI researches are focused on single-scale sampling. Despite trained with single-scale sampling, the learned single-scale matrix \cite{Shi17} mimics the conventional multi-scale sampling. That is, it captures more low-frequency components and results in the aliasing artifact of reconstructed images. Therefore, it would be easier for the network to learn multi-scale features with a multi-scale architecture.

Multi-scale DCI was introduced in \cite{Shi18, Shi19CVPR, Xu18} and our initial work \cite{Canh18B}. \textcolor{black}{LAPRAN} \cite{Xu18} defined a set of measurements correspond to a given resolution. The low-resolution image was recovered first and utilized to recover the higher resolution images. As a result, its low-resolution measurements capture more low-frequency components, thus follow the multi-scale prior. This method demands a heuristic measurement allocation for each resolution. \textcolor{black}{MS-CSNet \cite{Shi18}} trained a set of measurements with the corresponding subrate and reused at the larger subrate, therefore, their low-subrate measurements correspond to low-frequency components. \textcolor{black}{SCSNet \cite{Shi19CVPR} was a scalable framework that supports multiple levels of reconstructions at multiple levels of quality. Similar to LAPRAN, it favored the low-frequency contents more, especially at low frequency layers. However, SCSNet aimed to address the varying subrate problem (i.e. one network for multiple subrates) than designing a multi-scale DCI. Therefore, there is lacking rigorous study on multi-scale DCI}   

\begin{table}[!t]
	\renewcommand{\arraystretch}{1.3}
	\setlength{\tabcolsep}{4.5pt}	
	\caption{\textcolor{black}{Deep Compressive Imaging vs. Deep AutoEncoder}}
	\label{tab:DCIvsAuto}
	\centering
	\begin{tabular}{l|l|l}
		\hline \hline 
		Method & Deep Compressive Imaging & Deep AutoEncoder \\ 	\hline\hline
		Type   & End-to-End 			  & End-to-End \\ \cline{1-3}  
		\multirow{5}{0.13in}{Encoder}     & Less complexity  & Same complexity as decoder \\ 	\cline{2-3}  
	     								  & \multirow{2}{1.2in}{Large kernel size, i.e. $16\times 16, 32\times 32$} & \multirow{2}{1.2in}{Small kernel size, i.e. $3 \times 3, 5\times 5$} \\
										  & &\\ \cline{2-3}  
										  & Large stride, i.e. $16, 32$ & Small stride, i.e. $1, 2$ \\	 \cline{2-3}   
										  & Without padding & With padding \\ \cline{2-3}  
										  & \textcolor{black}{Without bias} & \textcolor{black}{With bias} \\ \hline							  		
		\multirow{4}{0.13in}{Decoder}     &	More complexity             & Same complexity as encoder 				\\ \cline{2-3}  					  
										  & Small kernel size, i.e. $3 \times 3$ 	& Small kernel size, i.e. $3 \times 3$  				\\ \cline{2-3} 
     									  & Small stride, i.e. $1, 2$ 	& Small stride, i.e. $1,2$ 				\\ \cline{2-3} 
     									  & With padding		        &  With padding				\\ 
	    \hline \hline
	\end{tabular}
\end{table}

\begin{figure*}[!t]
	\centering
	\label{fig:MSDCI}
	\includegraphics[scale=0.23]{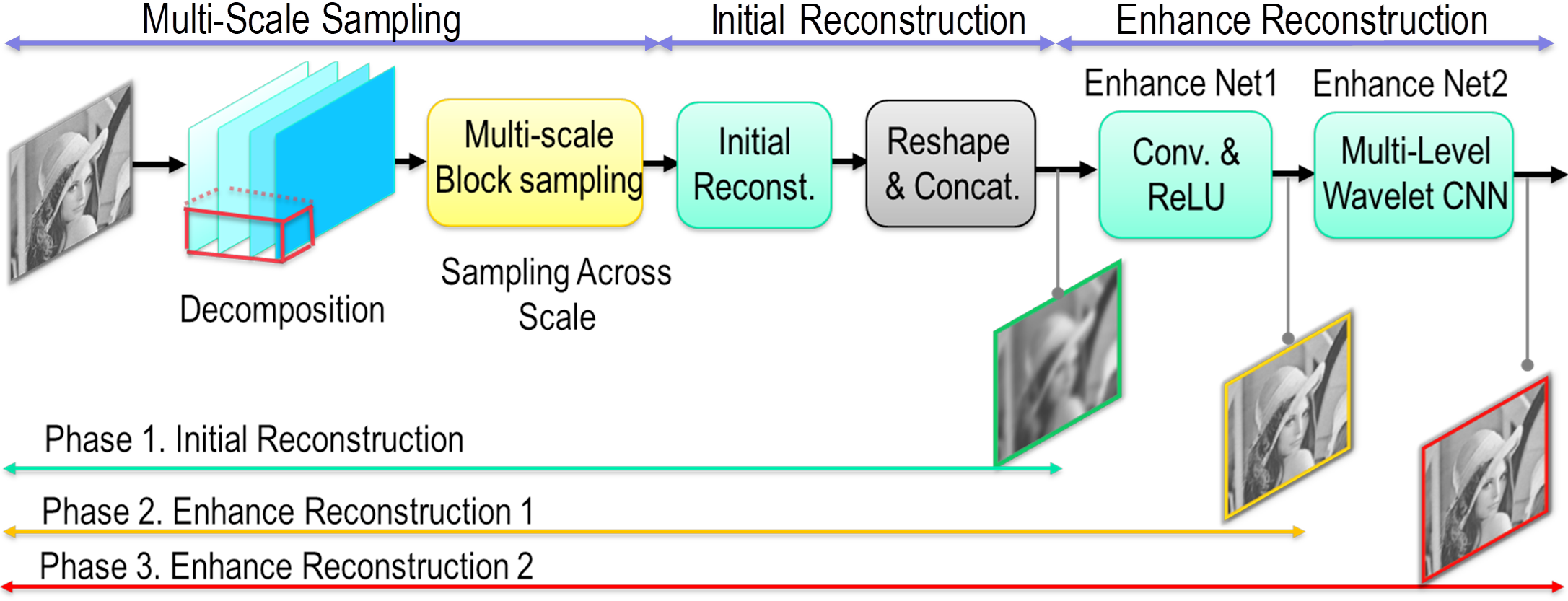}
	\caption{Multi-scale deep compressive imaging (MS-DCI) network with three-phase end-to-end training. Multi-scale sampling are detailed in Fig. \ref{fig:DecomMethods}. }
	\label{fig:MS_DCI}
\end{figure*}

\subsubsection{Deep Compressive Imaging vs. Deep AutoEncoder}
\textcolor{black}{Deep AutoEncoder (DAE) \cite{Mao16} is an unsupervised artificial neural network,} which learns the compact latent representation of signals similar to CS. We can interpret DCI as an asymmetric DAE with an encoder being much simpler than the decoder \textcolor{black}{(with an exception of \cite{Mousavi19} which designed equivalent complexity DCI)}. Also, DCI removes the non-linear activation and bias to maintain a simple encoder and compatible with the conventional CS. DCI uses a considerable large kernel size (e.g., $32\times 32$) while small kernel size is used in DAE (e.g. $3 \times 3$, $5\times 5$). Additionally, convolution in DCI is non-overlapping with the stride of convolution is equal to the sampling block size. \textcolor{black}{The summary is given in Table \ref{tab:DCIvsAuto}}. 

\textcolor{black}{Note that it is possible to implement non-linear activations and bias terms in the encoder (i.e., via post-processing) but only for one sampling layer to maintain practical. Thus, we study the impact of non-linearity and bias in Section V-B. }

\section{Multi-Scale Deep Compressive Imaging}

This section proposed a multi-scale deep compressive imaging (MS-DCI) which consists of three main parts of (i) multi-scale sampling, (ii) initial reconstruction, and (iii) multi-scale enhance reconstruction network as shown in Fig. \ref{fig:MS_DCI}. 

\begin{figure*}[!t]
	\centering
	\includegraphics[scale=0.43]{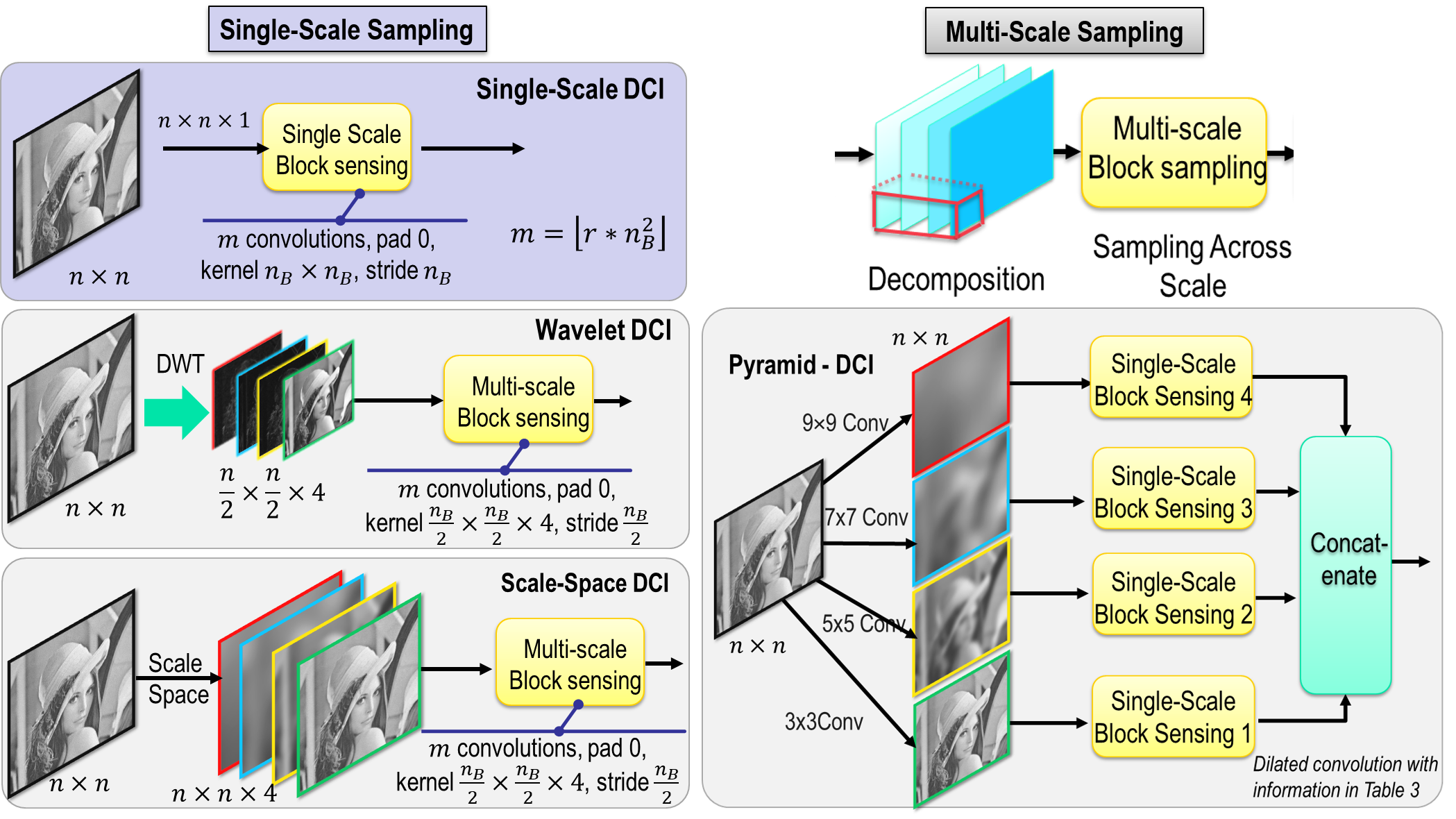}
	\caption{Single-scale (blue) vs multi-scale sampling (gray, with Wavelet, Scale-space, and Pyramid decomposition).  $n, n_B, m$ denote image size, block size, and number of measurements, respectively. Details on setting of P-DCI are given in Table \ref{tab:ConvParam_PDCI}.}
	\label{fig:DecomMethods}
\end{figure*}

\subsection{Multi-Scale Sampling Network}

\textcolor{black}{Conventional multi-scale sampling first linearly decomposes images into multiple scales (i.e. by wavelet decomposition), then adaptively acquires measurements at each signal's scale \cite{Fowler11, Canh18}. As the signal in each scale has a different level of sparsity, the number of measurements at different scale are heuristically allocated, thereby, limited the final performance.} 
We observe that single-scale sensing tends to prefer low-frequency contents similar to multi-scale sampling, especially at low-subrate. As shown in Fig. \ref{fig:SortFig}, the sorted energy of the learned matrix for single-scale CSNet is approximate the sorted energy of the multi-scale wavelet at low-low subband. Therefore, without additional information, CSNet struggles to learn multi-scale features and results in similar large variance kernels. Therefore, guiding multi-scale information is more likely to benefit the network to better capture multi-scale features. \textcolor{black}{Therefore, we design a multi-scale sampling architecture with multi-scale decomposition and multi-scale sampling which are modeled as convolution layers in a deep network to enable end-to-end learning.}

\textcolor{blue}{Firstly, a convolution layer without activation and bias is used to mimic the linear decomposition as 
\begin{equation}
    x_l = W_l^1  x, 
\end{equation}
where $W_l$, $l \in \{1, \dots ,c\}$ denotes the convolution kernel and decomposition filter, $x_l$ is decomposed signal at level $l$, and $c$ is the total number of level. It is straight forward to capture signal at each level independently by
\begin{equation}
    y_l = W_l^2  (W_l^1  x), = (W_l^2 W_l^1) x,  \quad y_l \in \mathbb{R}^{m_l}
\end{equation}
From this equation, we can acquire multi-scale measurements $y_{ms}$ with the multi-scale sampling matrix $\Phi_{ms}$
\begin{equation} 
\begin{split}
    y_{ms} &= \Phi_{ms} x,   \quad l \in \{1, 2, \cdots, c\} \\
    y_{ms} & = [y_0; y_1; \cdots; y_c], \quad \quad y_l = \Phi_l x, \\
    \Phi_{ms} & = [\Phi_0; \Phi_1; \cdots; \Phi_c], \quad \Phi_l = (W_l^2  W_l^1),  \\
\end{split}
\end{equation} 
However, this approach requires the number of measurement $m_l \in [0, m], \sum_l m_l = m$ at each level $l$ to allocate heuristically or adaptively. To overcome this drawback, we sample measurements across multiple decomposed scales as  
\begin{equation}
\begin{split}
    y_{ms}    &= \Phi_{ms} x, \quad  j \in [1, \cdots, m] \\
    y_j       &= \sum_l \left(W_l^{2, j}  W_l^{1}  x \right) = \sum_l \left(W_l^{2, j} W^1_l \right) x = \Phi_{ms}^j  x \\
    \Phi_{ms} &= [\Phi_{ms}^0, \cdots, \Phi_{ms}^m], \quad \Phi_{ms}^j = \sum_l \left(W_l^{2,j}  W_l^{1}\right),
\end{split}
\end{equation}
In both approaches, the multi-scale matrix is equivalent to a single sampling matrix $\Phi_{ms}$, thereby compatible with the sequential sampling scheme. By jointly learning to decompose and sensing at multi-scale, we can capture measurement more efficient. Additionally, after the training process, our multi-scale sampling matrices are fixed for all test images.}

\subsection{Multi-Phase Reconstruction Network}
\subsubsection{Initial Reconstruction }
Similar to \cite{Shi17}, this sub-network mimics the matrix inversion to recover multi-scale measurements. We can either (i) independently recover each scale then perform inverse decomposition to deliver the reconstructed image or (ii) directly recover image at the original domain. In our experiment, we select the later approach for better reconstruction quality. Similar to \cite{Shi17}, a reshape and concatenate layer is used after each $1 \times 1$ convolution to form the 2D recovered image. Our initial recovery is very simple with one convolution, without bias and activation. 

\subsubsection{Enhance Reconstruction }
The initial recovered image is subsequently enhanced by (i) simple convolutions and (ii) multi-scale wavelet convolutions \cite{Liu19}. The first enhanced network is a set of five convolutions (kernel size $3 \times 3$, stride 1, and zero padding 1) following by ReLU activation. We set the number of feature maps to 64 (except the first and last layer) which is identical to CSNet's architecture \cite{Shi17}. 
The second enhanced network uses multi-level wavelet convolution (MWCNN) \cite{Liu19} which applies convolutions on top of decomposed wavelet features to effectively capture features at multi-scale. Utilize MWCNN, our MS-DCI network can take advantage of multi-scale structures in both decomposition, sampling, and reconstruction. 

\subsection{Training}
\subsubsection{Loss Function }
Motivated by many image restoration methods \cite{Liu19, Zhang17, Shi17, Shi18, Shi19CVPR}, we select the Euclidean loss (L2 norm) as the objective function. We computed the average L2-norm error overall training samples as:
\begin{equation}
\text{min} \frac{1}{2N} \sum_{i=1}^N || f(x_i, \theta)-x_i ||_2^2 ,
\end{equation}
where $N$ is the total number of training samples, $x_i$ denotes an image sample, and $f$ is the network function at a setting $\theta$.  

\subsubsection{Multi-Phases Training }
As mentioned in \cite{Yao17}, the better the initial image is, the higher the quality reconstruction can be achieved. They first learned the initial reconstruction then enhanced the quality through a sub-network of residual architecture. Motivated by this approach, we propose a three-phase training process of (Phase 1) initial reconstruction, (Phase 2) enhance reconstruction by convolution, and (Phase 3) enhance reconstruction by MWCNN as visualized in Fig. \ref{fig:MS_DCI}. We train each phase sequentially from Phase 1 to Phase 3 follow the end-to-end setting. The previously trained network at $i$-th is used as initialization for the $(i+1)$-th phase. 

In each phase, the learning rate of 0.001, 0.0005, and 0.0001 are used for every 50 epochs. In Phase 3, we use the pre-trained MWCNN for Gaussian denoising at the noise level $15$ as initialization. The adaptive moment estimation (Adam) method is used for optimization. In general, only convolution, ReLU, and batch norm layers are used in our proposed deep compressive sensing network for reconstruction. 

\begin{table}[!t]
	\renewcommand{\arraystretch}{1.2}
	\caption{Convolution parameters for P-DCI at $4$ scales}
	\label{tab:ConvParam_PDCI}

	\centering
	\begin{tabular}{c|c|c|c|c}
		\hline \hline
		Dilate & Kernel size & Padding & \multirow{1}{0.35in}{Stride} & Block size \\ \hline 
		$d$   & $F_W \times F_H$ & $P_W \times P_H$ & $S$ & $n_B \times n_B$ \\\hline \hline 
		1	   & $32	\times 32$	     & $0\times	0$	& 32	& $32 \times 32$ 	\\ \hline
		2	   & $16	\times 16$	     & $0\times	0$	& 32	& $31\times31$ \\ \hline
		3	   & $11	\times 11$	     & $0\times	0$	& 32	& $31\times31$\\\hline
		4     & $9  \times 9 $ 	     & $1\times	1$   & 32	& $33\times33$\\\hline
		
		\hline \hline		
	\end{tabular}
\end{table}

\begin{table*}[!t]
	\centering	
	\renewcommand{\arraystretch}{1.2}
	\setlength{\tabcolsep}{4.5pt}	
	\caption{Performance comparison of various schemes in average PSNR [dB] and SSIM \textcolor{blue}{(first and second line of each row)}}
	\label{tab:Per01}
	\begin{tabular}{l c|cc"cc"cc|ccc|ccc|ccc}
		\hline \hline 		
		\multirow{2}{0.2in}{Image} & \multirow{2}{0.13in}{\scriptsize{Rate}} & \multirow{2}{0.22in}{MH}    & \multirow{2}{0.22in}{GSR}     & \multirow{2}{0.2in}{\scriptsize{DBCS}}  & \multirow{2}{0.2in}{\scriptsize{CSNet}} & \multirow{2}{0.2in}{\scriptsize{\textcolor{blue}{MS-CSNet}}} & \multirow{2}{0.22in}{\scriptsize{\textcolor{blue}{S-CSNet}}} &  \multicolumn{3}{c|}{\textcolor{blue}{P-DCI$^i$}}   & \multicolumn{3}{c|}{\textcolor{blue}{W-DCI$^i$}} & \multicolumn{3}{c}{\textcolor{blue}{SS-DCI$^i$}}       \\ \cline{9-17}
		&      &       &       &       &       &     &          												            & 1     & 2     & 3      & 1      & 2     & 3       & 1     & 2     & 3     \\  \hline 
		\multirow{6}{0.15in}{Set5}  & \multirow{2}{0.12in}{0.1}  & 28.57 & 29.98 &	 31.31 & 32.30 & 32.82    &  	32.77   & \textbf{30.84} & 32.59 & 33.27  & 30.66  & 32.44 & 33.39   & 30.50 & \textbf{32.69} &\textcolor{red}{\textbf{33.63}} \\
		&      													 & 0.821 & 0.865 &  0.894 & 0.902 & 0.909    &  	0.983   &\textbf{ 0.862}  & 0.906 & 0.914 & 0.855  & 0.904  & 0.917 & 0.863   &\textbf{ 0.941} & \textcolor{red}{\textbf{0.942}} \\ \cline{2-17}
		& \multirow{2}{0.12in}{0.2} 							 & 32.09 & 34.17 &   34.55 & 35.63 & 36.23    &  	36.15   & 34.04  & \textbf{35.94} & 36.28 & \textbf{34.06 } & 35.75  & 36.56 & 33.81   & \textbf{35.94} & \textcolor{red}{\textbf{36.68}} \\
		&     													 & 0.888 & 0.926 &	 0.940 & 0.945 & 0.949    &  	0.941   & 0.924  & 0.948 & 0.949 & \textbf{0.925}  & 0.941  & 0.951 & 0.924   & \textbf{0.949} & \textcolor{red}{\textbf{0.953}} \\ \cline{2-17}
		& \multirow{2}{0.12in}{0.3}  							 & 34.07 & 36.83 &   36.54 & 37.90 & 38.43    &  	38.45   & 36.11  & 37.95 & 38.42 & \textbf{36.51}  & 38.20  & 38.74 & 36.22   & \textbf{38.42} & \textcolor{red}{\textbf{38.92}} \\
		&      													 & 0.916 & 0.949 &   0.958 & 0.963 & 0.966    &  	0.963   & 0.948  & 0.963 & 0.964 & \textbf{0.952}  & 0.965  & 0.967 & 0.952   & \textbf{0.966 }& \textcolor{red}{\textbf{0.968}} \\ \hline
		\multirow{6}{0.15in}{Set14} & \multirow{2}{0.12in}{0.1}  & 26.38 & 27.50 &   28.54 & 28.91 & 29.29    &  	29.22   &\textbf{ 27.87}  & 29.15 & 29.56 & 27.81  & 29.10  & 29.67 & 27.69   & \textbf{29.22} & \textcolor{red}{\textbf{29.69}} \\
		&     													 & 0.728 & 0.771 &   0.832 & 0.812 & 0.820    &  	0.818   & \textbf{0.783}  & 0.816 & 0.827 & 0.778  & 0.815  & 0.828 & 0.784   & \textbf{0.823} & \textcolor{red}{\textbf{0.837}} \\ \cline{2-17}
		& \multirow{2}{0.12in}{0.2}  							 & 29.47 & 31.22 &   31.21 & 31.86 & 32.26    &  	32.19   & 30.63  & 32.13 & 32.43 & \textbf{30.69}  & 32.05  & 32.51 & 30.43   & \textbf{32.23} & \textcolor{red}{\textbf{32.72}} \\
		&      													 & 0.824 & 0.864 &   0.890 & 0.891 & 0.896    &  	0.824   & 0.873  & 0.893 & 0.896 & \textbf{0.874}  & 0.893  & 0.900 & 0.872   & \textbf{0.898} & \textcolor{red}{\textbf{0.904}} \\ \cline{2-17}
		& \multirow{2}{0.12in}{0.3}  							 & 31.37 & 33.74 &   33.08 & 33.99 & 34.34    &  	34.51   & 32.52  & 34.05 & 34.32 & \textbf{32.86}  & 34.30  & 34.71 & 32.63   & \textbf{34.50} & \textcolor{red}{\textbf{34.90}}  \\
		&     													 & 0.869 & 0.907 &   0.921 & 0.928 & 0.919    &  	0.928   & 0.912  & 0.927 & 0.929 & \textbf{0.917}  & 0.930  & 0.934 & 0.917   & \textbf{0.933} & \textcolor{red}{\textbf{0.937}} \\ \hline \hline 
		\multicolumn{17}{r}{\scriptsize{The best results in each training phase $i$ of MS-DCI and all methods are in \textbf{Bold} and {\textcolor{red}{red}}}, respectively. Multi-scale methods are in \textcolor{blue}{blue}} \\ 		
	\end{tabular}
\end{table*}

\section{Proposed Decomposition Methods for MS-DCI}
\subsection{\textcolor{black}{Linear Image Decomposition}}

In general, images can be decomposed into multiple layers each carries a different amount of signal information for image analysis. We aim to study linear image decompositions as then can be easily integrated into the linear projection framework of CS \cite{Canh18}. Without loss of generality, we linearly decompose an image $x \in \mathbb{R}^N$ into $c$ layers, $\{x_l\}_{l=1}^c$ by a Pyramid decomposition formula as
\begin{equation}
x_l= W_l \cdot x = D_l \cdot  B_l \cdot x, \quad l \in \{1, \dots ,c\} ,
\end{equation}
where $D_l$ denotes a down-sampling matrix, and $B_l$ represent a smoothing matrix at layer $l$-th.  It is possible to derived various linear decomposition model from eq. (3) such as: (i) Scale-space decomposition by setting $D_l$ to an identity matrix; (ii) multi-resolution by setting $B_l$ to an identity matrix; (iii) wavelet decomposition by using high and low-pass filters $B_l$. 

\subsection{\textcolor{black}{Multi-Scale DCI with Decomposition}}
This section realizes various decomposition methods in a deep multi-scale sampling network, with corresponding multi-scale sampling sub-network. Details for parameters of all frameworks are presented in Fig. \ref{fig:DecomMethods}. 

\subsubsection{Wavelet-based DCI  (W-DCI)}
We implement the Discrete Haar Wavelet Transform (DWT) transform as a layer in a deep network. Given an image of size $n \times n$, DWT outputs four frequency bands at size $\frac{n}{2}\times \frac{n}{2} \times 4$. 
Then, we perform multi-scale block sampling with $m_i$ convolutions across decomposed channels with kernel size $\frac{n_B}{2} \times \frac{n_B}{2} \times 4$, without bias and activation [14]. Notation $n_B$ denotes the block size, and $m_i$ stands for the number of measurements for each block at a given sampling rate $r= m_i/n^2_B$.  

\subsubsection{Scale-space DCI (SS-DCI)}
In the Scale-space analysis [35], a signal is decomposed into multiple layers using a set of Gaussian smoothing filters. To integrate Scale-space to deep network, we model the smoothing process as convolution so that the decomposition also can be trained (unlike fixed decomposition kernel in W-DCI). To comparable with W-DCI, we use four convolutions with the different kernel: $3 \times3, 5 \times5, 7 \times7$, and $9 \times 9$ to output four decomposed features. Multi-scale sampling is performed with $m_i$ convolutions with kernel size $\frac{n_B}{2} \times \frac{n_B}{2} \times 4$, without bias and activation. Note that, Scale-space decomposition shares similarity with GoogleNet [36] which exploits multiple filter kernel size. 

\subsubsection{Pyramid-based DCI (P-DCI)} 
The Pyramid decomposition [37] is equivalent to Scale-space decomposition with additional down-sampling operators. Avoid using two operations for downscale and sampling, we use dilated convolutions \cite{Yu16} instead. Dilated convolution is a subsampled convolution with the fixed sampling grid. It is equivalent to down-scale image first (by the nearest neighbor down-sampling) and follow by a conventional convolution. Unlike W-DCI and SS-DCI, P-DCI captures images at each scale independently due to the difference in resolution (or dilated factor). The details settings for P-DCI are given in Table \ref{tab:ConvParam_PDCI}.   

\section{Experimental Results}
For experiments, we used the DIV2K dataset \cite{Timofte17} to generate training data with $64 \times 500$ patches of grayscale images of size $256 \times 256$. Our MS-DCIs are implemented under the MatConvNet \cite{Vedaldi15}. 
\textcolor{black}{Table \ref{tab:Algorithms} summaries the related work for comparison. For a fair comparison, we selected block size $32 \times 32$ for single-scale and $16\times16$ for multiple-scale sensing except for P-DCI as in Table \ref{tab:ParamCompare}. We defined MS-DCI at different training phases as W-DCI$^i$, SS-DCI$^i$, and P-DCI$^i$ with $i=1,2,3$. Set5, Set14 \cite{Timofte17} and test images of $512 \times 512$ are evaluated.} 

\begin{table}
	\centering	
	\renewcommand{\arraystretch}{1.2}
	\setlength{\tabcolsep}{5pt}	
	\caption{List of sampling methods for comparison}
		\label{tab:Algorithms}
	\begin{tabular}{lll}
		\hline 
		\multirow{2}{2.0cm}{Method}               & \multicolumn{2}{c}{Sampling} \\ \cline{2-3}
		& \multirow{1}{*}{Single-scale}  	& \multirow{1}{2cm}{Multi-scale}  \\ \hline
		\multirow{3}{*}{Conventional} & MH \cite{Chen11}        &    MRKCS \cite{Canh15}     \\
		& GSR \cite{Zhang14}        &      \\
		& RSRM \cite{Canh19}        &            \\	\hline 
		\multirow{5}{*}{Deep Learning}       	& ReconNet \cite{Kul16}        & MS-CSNet \cite{Shi18}    \\
		&  DR$^2$Net  \cite{Yao17}   & SCSNet  \cite{Shi19CVPR}      \\
		&  DBCS  \cite{Adler17}     & \textbf{P-DCI$^{1,2,3}$}        \\
		& CSNet \cite{Shi17}        & \textbf{W-DCI$^{1,2,3}$}        \\
		& KCSNet  \cite{Canh18A}    & \textbf{SS-DCI$^{1,2,3}$ }     \\ \hline 
		\multicolumn{3}{r}{\scriptsize{Proposed methods are highlighted in \textbf{Bold}.}}
	\end{tabular}
\end{table}

\begin{table}[]
	\renewcommand{\arraystretch}{1.05}
	\setlength{\tabcolsep}{3.5pt}	
	\caption{Comparison of various decomposition schemes }
		\label{tab:ParamCompare}

	\centering
	\begin{tabular}{c |c|cc|c|ccc}
		\hline \hline 
		\multirow{3}{*}{Method} & \multirow{3}{0.5in}{Decomp. Image Size}        & \multicolumn{2}{c|}{Decomp. Filter}          & \multicolumn{4}{c}{Sampling Matrix}                                        \\ \cline{3-8}
		&                                            & \multirow{2}{0.1in}{Size} & \multirow{2}{0.1in}{No.} & \multirow{2}{*}{kernel size} & \multicolumn{3}{c}{No. meas. /block} \\ \cline{6-8}
		&                                            &                       &                      &                                    & 0.1         & 0.2        & 0.3        \\ \hline \hline 
		CSNet                   & $n\times n \times 1$                       & -                     & -                    & $32 \times 32 \times 1$            & 102         & 204        & 307        \\ \hline
		W-DCI                   & $\frac{n}{2}\times \frac{n}{2} \times 1$ & $2\times 2$           & 4                    & $16\times 16 \times 4$             & 102         & 204        & 307        \\ \hline
		\multirow{4}{*}{SS-DCI} & \multirow{4}{*}{$n\times n \times 4$}      & $3\times 3$           & 1                    & $16 \times 16 \times 1$            & \multirow{4}{*}{26}           & \multirow{4}{*}{51}          & \multirow{4}{*}{102}         \\ \cline{3-5}
		&                                            & $5\times 5$           & 1                    & $16 \times 16 \times 1$            &    &      &        \\  \cline{3-5}
		&                                            & $7\times 7$           & 1                    & $16 \times 16 \times 1$            &           &          &         \\  \cline{3-5}
		&                                            & $9\times 9$           & 1                    & $16 \times 16 \times 1$            &           &          &         \\ \hline
		\multirow{4}{*}{P-DCI}  & \multirow{4}{*}{$n\times n \times 4$}      & $3\times 3$           & 1                    & $32 \times 32 \times 1$            & 26          & 51         & 102        \\ \cline{3-8}
		&                                            & $5\times 5$           & 1                    & $16 \times 16 \times 1$            & 26          & 51         & 102        \\ \cline{3-8}
		&                                            & $7\times 7$           & 1                    & $11 \times 11 \times 1$            & 26          & 51         & 102        \\ \cline{3-8}
		&                                            & $9\times 9$           & 1                    & $9 \times 9 \times 1$              & 24          & 52         & 101    \\  \hline \hline 
	\end{tabular}
\end{table}

\begin{figure*}[!t]
	\centering
	\includegraphics[scale=0.33]{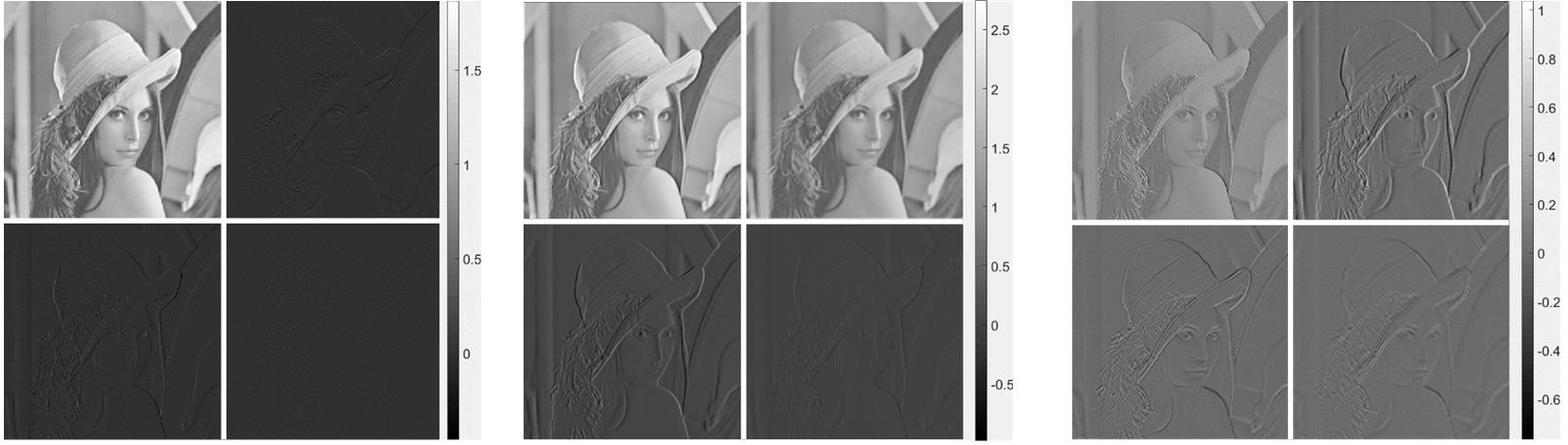}
	\caption{Decomposed image with various MS-DCI: (left to right) W-DCI$^3$, P-DCI$^3$, and SS-DCI$^3$. The vertical axis indicates the range of features. The gradually change the decomposed range between scales in W-DCI$^3$ and $P-DCI^3$ indicates that networks can learn to decompose image to multi-scale.}
	\label{fig:decomp_imgs}
\end{figure*}

\begin{figure*}[!t]
	\centering
	\includegraphics[scale=0.33]{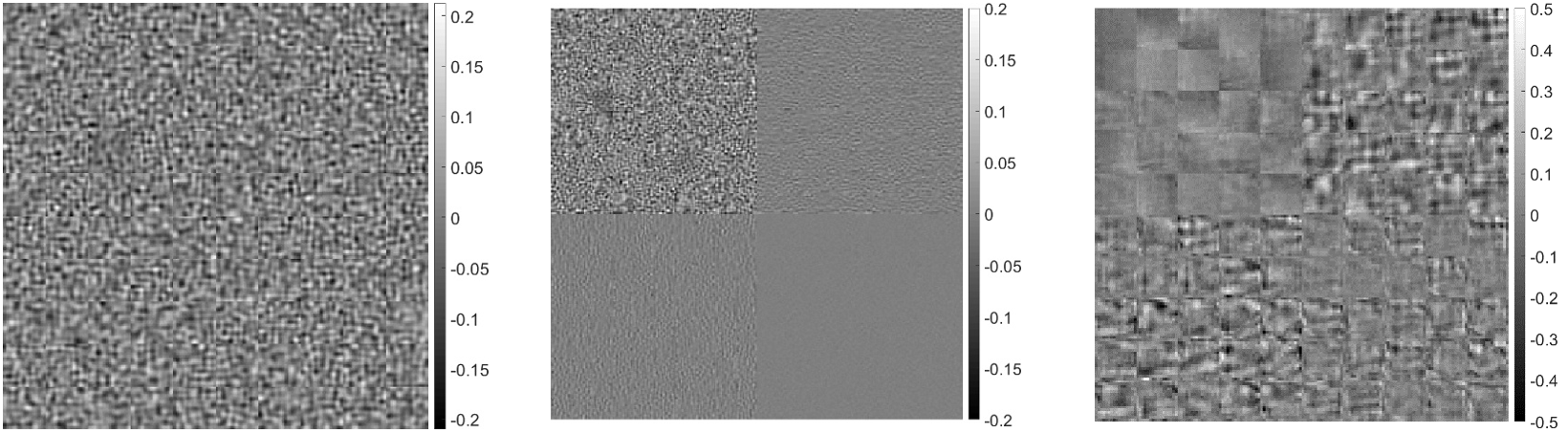}
	\caption{Learned measurement matrix of (left to right) CSNet, W-DCI$^3$, and SS-DCI$^3$ at subrate 0.1. The vertical axis indicates the range of features. The significant difference in different in measurements at each scale indicate that network can learn to sample image at multi-scale.}
	\label{fig:learn_matrix}
\end{figure*}

\subsection{Multi-Phase Training Performance }

The efficiency of multiple phase training was examined by the reconstruction quality in Table \ref{tab:Per01}. 
The best reconstruction quality in Phase 1 was P-DCI$^1$ and W-DCI$^1$ at low (0.1) high subrate (0.2 and 0.3), respectively. Better capturing the low-frequency components is the main reason with higher dilated convolution for the low-frequency band in P-DCI$^1$ and well separate low and high-frequency of DWT in W-DCI$^1$. 
At Phase 2, with better reconstruction, learned decomposition in SS-DCI$^2$ and P-DCI$^2$ offered higher PSNR than fixed decomposition kernel of W-DCI$^2$. Because of multi-scale sampling, all multi-scale methods showed $0.14 \sim 0.31$dB gain and similar reconstruction quality as single-scale (CSNet) and multi-scale DCIs (MS-CSNet and S-CSNet). 

At Phase 3, we utilized the multi-scale reconstruction with MWCNN with significant additional parameters and complexity, thereby, greatly improved the final reconstruction performance. The order of increasing reconstruction quality was P-DCI$^3$, W-DCI$^3$, and SS-DCI$^3$. Thanks to learned decomposition, SS-DCI$^3$ showed $0.17 \sim 0.28$dB improvement over W-DCI$^3$. On the other hand, our P-DCI$^3$ presented the poorest reconstruction quality due to favoring to low-frequency components and independent sampling each decomposed scale. 

What stands out from Table \ref{tab:Per01} was the increase of reconstruction quality along with the training phase. Compared to Phase 1, Phase 2 and Phase 3 improved $0.70\sim 0.12$dB (Set5) and $1.21 \sim 1.70$dB (Set14) in average, respectively. From Table \ref{tab:params_vars}, Phase 2 and Phase 3 added 0.21 and 16.17 million more parameters to the reconstruction. In general, more parameters is, higher learning capability, and results in better reconstruction performance.

\begin{figure*}[!t]
	\centering
	\includegraphics[scale=0.33]{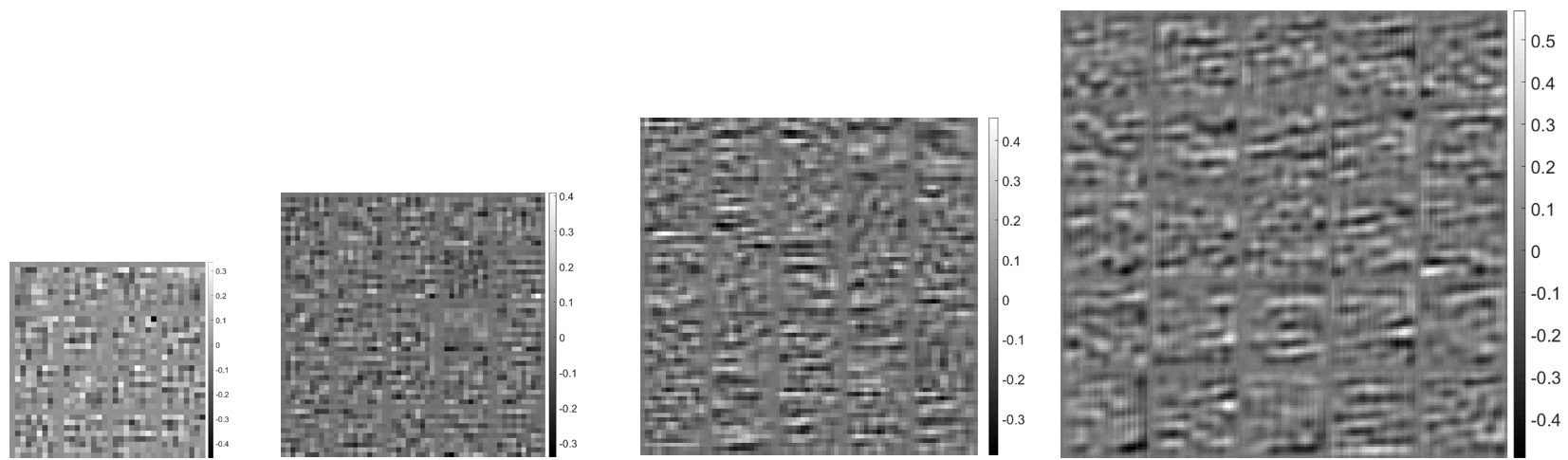}
	\caption{Learned measurement matrix of P-DCI$^3$ at subrate 0.1 at kernel size: (left to right) $9\times 9, 11\times11, 16\times16$, and $32 \times 32$. The vertical axis indicates the range of features. }
	\label{fig:learn_matrix_PDCI}
\end{figure*}

\begin{figure*}[!t]
	\centering
	\includegraphics[scale=0.33]{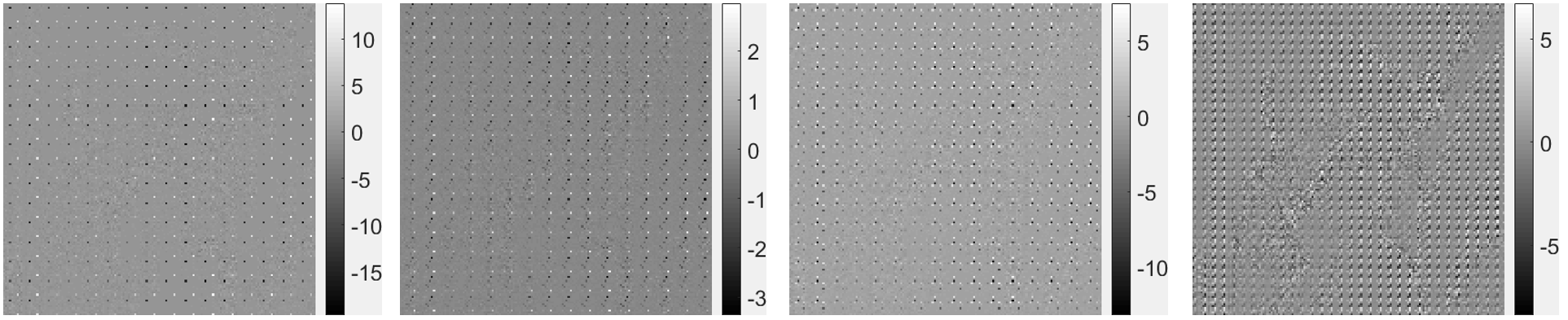}
	\caption{Visualization of the learned measurements for Lena image at subrate 0.1: (left to right) CSNet, W-DCI$^3$, P-DCI$^3$, and SS-DCI$^3$. Vertical axis indicates the range of learned measurements.}
	\label{fig:learn_meas}
\end{figure*}

\subsection{Evaluate Decomposition Methods }

\subsubsection{\textcolor{black}{Reconstruction Performance}}

\textcolor{black}{The number of parameters does not linearly proportional to the learning capability. While multi-scale sampling accounts for less than $1\%$ of parameters in Phase 3, it still plays a significant role. At slightly smaller parameters, SS-DCI$^3$ still showed better performance with up to $0.24$dB than W-DCI$^3$ and $0.58$ over P-DCI$^3$, respectively. This difference is similar to Phase 2. It is because SS-DCI learned a better decomposition than the fixed wavelet decomposition with DWT. Also, decomposed images of SS-DCI are at the higher resolution $n\times n$ than $\frac{n}{2} \times \frac{n}{2}$ in W-DCI and P-DCI, thus, easier to capture the high-frequency contents. }

\begin{figure}[!t]
	\centering
	\includegraphics[scale=0.2]{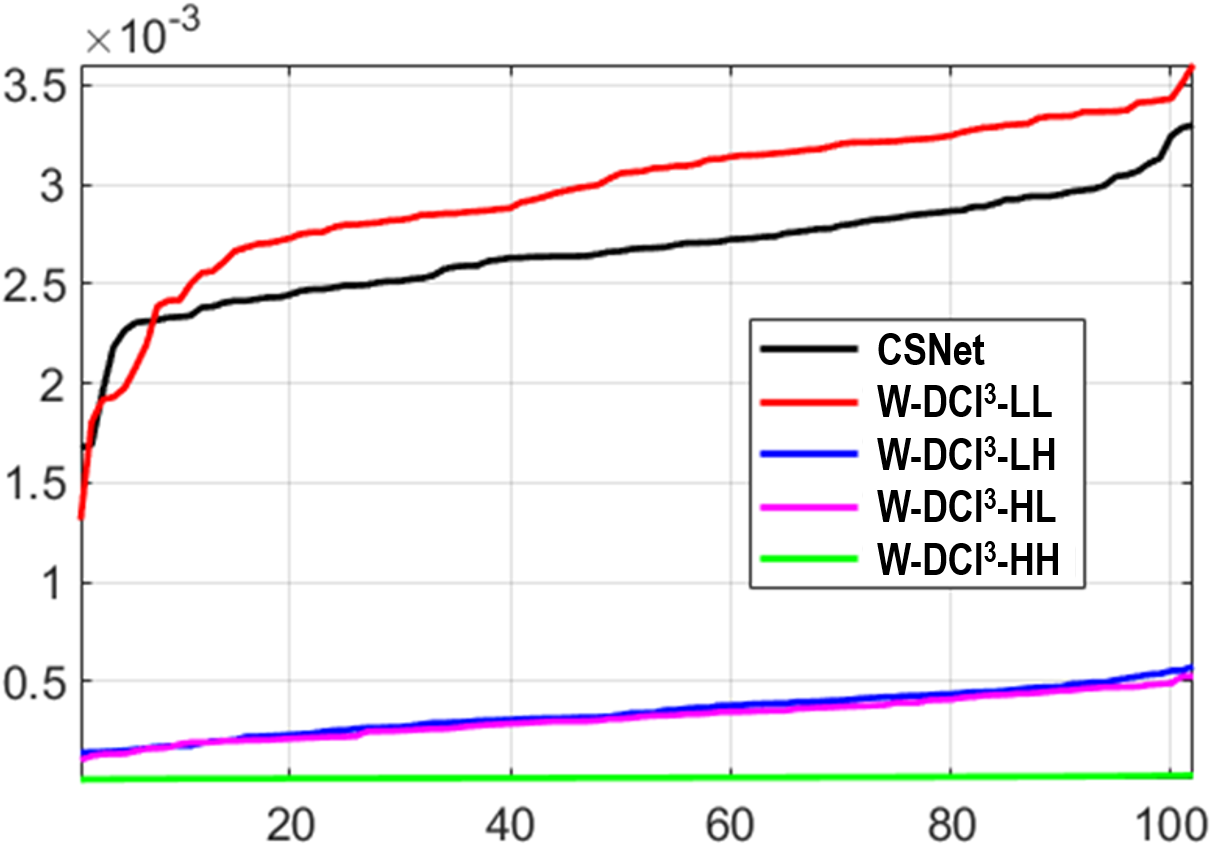}
	\caption{\textcolor{black}{Sorted variance of learned kernels in CSNet and W-DCI$^3$. There are significant different in variance between low-frequency (low-low) and high frequency bands (low-high, high-low, and high-high) in W-DCI$^3$.}}
		\label{fig:SortFig}
\end{figure}

\subsubsection{Learned Decomposition}
We showed  four decomposed images for our MS-DCIs in Fig. \ref{fig:decomp_imgs}. It is clearly shown that P-DCI and SS-DCI network can learn to decompose the image with gradually reduce features ranges between scales. 

Note that, the size of decomposed images W-DCI are $\frac{n}{2} \times \frac{n}{2}$ in W-DCI and $n \times n$ in both P-DCI and W-DCI. With Haar DWT, W-DCI$^3$ divided images into four distinct subbands with a significant difference in energy. Meanwhile, SS-DCI has departed from the conventional scale-space (i.e., each scale is a Gaussian smoothed image) and somehow mimicked the decomposition of W-DCI. That is, the first level looks like the low-low (i.e., image content is visible) and the other scales look similar to the high-frequency parts. The intensity is also gradually reduced along with the scale. On the other hand, in P-DCI$^3$, we observed a similar pattern to SS-DCI$^3$ but less distinctive between scales. Also, the difference between scales in SS-DCI and P-DCI are not as substantial as W-DCI.

\begin{table*}[]
	\label{tab:PerSOTA}
	\renewcommand{\arraystretch}{1.05}
	\setlength{\tabcolsep}{3.5pt}	
	\caption{Reconstruction performance of various sampling methods in PSNR [dB] and SSIM (first and second line of every row).}
	\centering
	\begin{tabular}{lc" ccc" cccc|c" ccc}
		\hline \hline 
		\multirow{1}{0.3in}{Image}                  &  rate & \space \space  GSR \space  & \scriptsize \textcolor{blue}{MRKCS} & \scriptsize RSRM  & \scriptsize ReconNet & \scriptsize DR$^2$Net & \scriptsize \space CSNet \space & \scriptsize KCSNet  & \scriptsize \textcolor{blue}{S-CSNet} & \scriptsize \textcolor{blue}{P-DCI$^3$} & \scriptsize \textcolor{blue}{W-DCI$^3$} & \scriptsize \textcolor{blue}{SS-DCI$^3$} \\ \hline \hline 
		\multirow{6}{*}{Lena}  & \multirow{2}{*}{0.1} & 30.97 & 32.87 & \textbf{33.31} & 26.89    & 28.65  & 32.15 & 33.09  			  &	32.36	& 32.63  & 32.90   & 33.05   \\
		&                      & 0.866 & 0.821 & 0.887 & 0.749    & 0.800    & 0.879 & 0.887   					   		 	  &	0.882	& 0.887  & 0.891  & \textbf{0.895 }  \\ \cline{2-13}
		& \multirow{2}{*}{0.2} & 34.44 & 35.85 & \textbf{36.41} & \multirow{2}{*}{-} & \multirow{2}{*}{-}  & 35.21 & 35.08  		 	  & 35.41 	& 35.75  & 35.86  & {36.14}   \\ 
		&                      & 0.914 & 0.919 & 0.925 &          &        & 0.924 & 0.913  					   		 	  & 0.910 	& 0.926  & 0.929  & \textbf{0.931}   \\ \cline{2-13}
		& \multirow{2}{*}{0.3}&36.47& 37.72 & 38.00 &\multirow{2}{*}{-} &\multirow{2}{*}{-} &37.33 &\multirow{2}{*}{-}   	  & 37.51	& 37.44  & 37.85  & \textbf{38.11 }  \\
		&                      & 0.936 & 0.940  & 0.941 &          &        & 0.945 &        					   		      & 0.946   & 0.944  & 0.948  & \textbf{0.950}    \\  \hline
		\multirow{6}{0.3in}{Peppers}  & \multirow{2}{*}{0.1} & 31.45 & 33.40  & 32.42 & 26.21 & 28.32  & 32.06 & 31.00        & 32.57 	& 33.17  & 33.48  & \textbf{33.52}   \\
		&                      & 0.843 & 0.853 & 0.845 & 0.716    & 0.769  & 0.858 & 0.861  							 	  & 0.863 	& 0.869  & 0.87   & \textbf{0.876}   \\\cline{2-13}
		& \multirow{2}{*}{0.2} & 34.12 & 35.52 & 34.89 & \multirow{2}{*}{-}    & \multirow{2}{*}{-} & 34.42 & 32.80      	  &	35.05   & 35.32  & 35.47  & \textbf{35.58}   \\
		&                      & 0.88  & 0.891 & 0.881 &          &        & 0.891 & 0.881 									  &	0.896	& 0.897  & 0.899  & \textbf{0.901}   \\\cline{2-13}
		& \multirow{2}{*}{0.3} &35.65& 36.59 & 36.31 & \multirow{2}{*}{-} &\multirow{2}{*}{-}&35.84 &\multirow{2}{*}{-}  	  & 36.49   & 36.42  & 36.68  & \textbf{36.74}   \\
		&                      & 0.905 & 0.909 & 0.905 &          &        & 0.910  &        								  &	0.915   & 0.913  & 0.915  & \textbf{ 0.918}   \\ \hline
		\multirow{6}{0.3in}{Mand-rill} & \multirow{2}{*}{0.1} & 19.93 & 21.92 & 20.12 & 19.70  & 20.18  & 22.26& 22.18   	  & 22.29	& 22.48  & 22.5   & \textbf{22.62}   \\
		&                      & 0.508 & 0.549 & 0.491 & 0.411    & 0.455  & 0.592 & 0.581   							 	  &	0.597	& 0.611  & 0.610  & \textbf{0.624}   \\\cline{2-13}
		& \multirow{2}{*}{0.2} & 22.22 & 23.61 & 22.52 & \multirow{2}{*}{-}        & \multirow{2}{*}{-} & 24.08 & 23.51  	  &	24.19	& 24.40  & 24.44  & \textbf{24.57}   \\
		&                      & 0.682 & 0.688 & 0.659 &          &        & 0.749 & 0.697 								 	  &	0.754	& 0.762  & 0.767  & \textbf{0.775}   \\\cline{2-13}
		& \multirow{2}{*}{0.3} &23.92 & 25.13  &24.40&\multirow{2}{*}{-}&\multirow{2}{*}{-}& 25.72 & \multirow{2}{*}{-}  	  &	25.88	& 25.89  & 26.05  & \textbf{26.31  } \\
		&                      & 0.775 & 0.780 & 0.751 &          &        & 0.833 &        								  &	0.838	& 0.838  & 0.842  &\textbf{ 0.854}   \\\hline
		\multirow{6}{*}{Boats}     & \multirow{2}{*}{0.1} & 27.55 & 28.78 & 27.82 & 24.35    & 24.35  & 29.08 & 28.99   	  &	29.41 	& 29.72  & 29.66  & \textbf{29.87}   \\
		&                      & 0.773 & 0.786 & 0.762 & 0.636    & 0.636  & 0.812 & 0.802   							     & 0.822 	& 0.833  & 0.833  & \textbf{0.841}   \\\cline{2-13}
		& \multirow{2}{*}{0.2} & 31.34 & 31.88 & 32.00 & \multirow{2}{*}{-}  & \multirow{2}{*}{-}   & 32.05 & 30.93      	  & 32.47 	& 32.76  & 32.82  & \textbf{33.10}    \\
		&                      & 0.862 & 0.865 & 0.865 &          &        & 0.884 & 0.852  							 	  & 0.891 	& 0.894  & 0.896  & \textbf{0.899}   \\\cline{2-13}
		& \multirow{2}{*}{0.3} & 33.72 & 33.73 & 34.09 & \multirow{2}{*}{-}   & \multirow{2}{*}{-}   & 33.98 & -      	 	  & 34.41 	& 34.29  & 34.73  & \textbf{35.05}   \\
		&                      & 0.904 & 0.899 & 0.901 &          &        & 0.911 &         								  & 0.916 	& 0.914  & 0.919  & \textbf{0.922}   \\\hline
		\multirow{6}{0.4in}{Camera-man} & \multirow{2}{*}{0.1} & 32.12 & \textbf{34.51} & 34.44 & 26.03 & 28.46 & 31.15 & 32.98   	  & 31.36   & 33.46  & 33.10  & 33.56   \\
		&                      & 0.913 & 0.928 & 0.927 & 0.798    & 0.848  & 0.918 & 0.930    							 	  & 0.925   & 0.940  & 0.942  & \textbf{0.945}   \\\cline{2-13}
		& \multirow{2}{*}{0.2} & 37.15 & 38.92 & 36.57 & \multirow{2}{*}{-}   & \multirow{2}{*}{-}  & 34.59 & 36.37      	  & 36.82 	& 38.95  & 39.46  & \textbf{39.68}   \\
		&                      & 0.958 & 0.967 & 0.969 &          &        & 0.961 & 0.963   							 	  & 0.975 	& 0.977  & 0.980   & \textbf{0.982}   \\\cline{2-13}
		& \multirow{2}{*}{0.3}&40.58& 42.37&42.54 & \multirow{2}{*}{-} &\multirow{2}{*}{-} & 37.47 &\multirow{2}{*}{-}   	  & 40.92 	& 42.84  & 44.10   & \textbf{44.27}   \\
		&                      & 0.977 & 0.984 & 0.981 &          &        & 0.976 &         							 	  & 0.990 	& 0.989  & 0.992  & \textbf{0.993}   \\\hline
		\multirow{6}{*}{Man}       & \multirow{2}{*}{0.1} & 27.74 & 29.58 & 28.21 & 25.30     & 26.51  & 29.84 & 29.86   	  & 30.03   & 30.17  & 30.21  & \textbf{30.44}   \\
		&                      & 0.781 & 0.812 & 0.774 & 0.660     & 0.714  & 0.833 & 0.829   							 	  & 0.840   & 0.845  & 0.847  & \textbf{0.855}   \\\cline{2-13}
		& \multirow{2}{*}{0.2} & 30.63 & 32.31 & 32.39 & \multirow{2}{*}{-} & \multirow{2}{*}{-} & 32.55 & 31.81 		 	  & 32.77 	& 32.87  & 32.87  & \textbf{33.05}   \\
		&                      & 0.867 & 0.885 & 0.885 &          &        & 0.907 & 0.883   							 	  & 0.915 	& 0.912  & 0.914  & \textbf{0.917}   \\\cline{2-13}
		& \multirow{2}{*}{0.3} & 32.83 & 34.33 & 34.94 &\multirow{2}{*}{-}&\multirow{2}{*}{-} &34.52&\multirow{2}{*}{-}  	  & 34.76 	& 34.55  & 34.94  & \textbf{35.06}   \\
		&                      & 0.921 & 0.924 & 0.925 &          &        & 0.939 &         							 	  & 0.942 	& 0.939  & 0.944  & \textbf{0.946}   \\ \hline \hline 
		\multirow{6}{*}{Average}      & \multirow{2}{*}{0.1} & 28.29 & 30.18 & 29.39 & 24.75  & 26.29  & 29.42 & 29.68   	  & 29.97 	& 30.27  & 30.31  & \textbf{30.51}   \\
		&                      & 0.781 & 0.792 & 0.792 & 0.662    & 0.712  & 0.815 & 0.815  							 	  & 0.817 	& 0.831  & 0.832  & \textbf{0.839}   \\\cline{2-13}
		& \multirow{2}{*}{0.2} & 31.65 & 33.02 & 32.46 & \multirow{2}{*}{-}  &  \multirow{2}{*}{-}  & 32.15 & 31.75      	  & 32.79 	& 33.34  & 33.49  & \textbf{33.69}   \\
		&                      & 0.861 & 0.869 & 0.864 &          &        & 0.886 & 0.865 								 	  & 0.890 	& 0.895  & 0.898  & \textbf{0.901}   \\\cline{2-13}
		& \multirow{2}{*}{0.3} & 33.86 & 34.98&35.05&\multirow{2}{*}{-}&\multirow{2}{*}{-} & 34.14 &\multirow{2}{*}{-}   	  & 35.00 	& 35.24  & 35.73  & \textbf{35.92}   \\
		&                      & 0.903 & 0.906 & 0.901 &          &        & 0.919 &        							 	  & 0.925 	& 0.923  & 0.927  & \textbf{0.931}  \\\hline \hline 
		\multicolumn{13}{r}{Multi-scale methods are in \textcolor{blue}{blue} and the best performance is in \textbf{Bold}.}  
	\end{tabular}
\end{table*}

\subsubsection{Learned Multi-Scale Sampling Matrices }
We visualized the learned measurement matrices in Fig. \ref{fig:learn_matrix}. We selected part of the learned matrix, reshaped, and concatenated them to make a larger image for better visibility.  For CSNet, we selected 100 over 102 filter at block size $32 \times 32$ and sub-rate 0.1. In general, it is difficult to observe any pattern in the learned kernel of CSNet. Single scale sampling like CSNet has learned to capture both low and high-frequency contents and results in similar variance of learned kernels in Fig. \ref{fig:SortFig}.    

For W-DCI, we visualized (over 102) sampling block size of $16 \times 16 \times 4$ at sub-rate 0.1, four Wavelet scales. At each scale, the learned kernels were concatenated to form a larger image for better visualization. \textcolor{black}{The horizontal, vertical, as well as diagonal patterns of the learned sampling matrix, were correspond to the wavelet band of low-high (LH), high low (HL), and high-high (HH), respectively. In Fig. \ref{fig:SortFig}, unlike the single scale CSNet, there was a significant difference in the variance of learned kernels between low-low (LL) and other bands in W-DCI$^3$, thus verified the effectiveness of multi-scale sampling.} 

Instead of reducing the dimension of decomposed images, SS-DCI reduced the total number of filters. There are 26 filters of size $16 \times 16 \times 4$ at subrate 0.1. We showed only 25 filters that were reshaped and concatenated in Fig. \ref{fig:learn_matrix}. Similarly, we illustrated the learned sampling matrices of P-DCI$^3$ with 25 filters for each scale with the corresponding size $9 \times 9, 11\times 11, 16\times 16$, and $32\times 32$ in Fig. \ref{fig:learn_matrix_PDCI}. 
It is shown that the network can learn to efficiently sampling at multi-scale as distinctive variance at different scales. The learned sampling matrices of both P-DCI$^3$ and SS-DCI$^3$ at the low-frequency measurements are smoother in comparison with that of the high-frequency, but not as significant different as the learned sampling matrices of W-DCI$^3$.

\subsubsection{Learned Multi-Scale Measurements}

We visualized the learned measurements in Fig. \ref{fig:learn_meas}. The measurement size of CSNet was $32\times 32\times 102$ at subrate 0.1, we took 100 measurements, reshaped to $10 \times 10$, and concatenated them to form a larger measurement image. Similarly, we selected $16\times 16\times 100$ for W-DCI$^3$. On the other hand, the measurement for SS-DCI$^3$ and P-DCI$^3$ were in $32\times 32\times 26$ at subrate 0.1. As a result, we selected only $25$ measurements to form a larger image for better visualization. 

It is easy to observe that SS-DCI$^3$ captured edges features more efficient with the stronger visible structures in its captured measurements. In contrast, it also revealed more information about the sampled images. As visualized in Fig. \ref{fig:learn_meas}, the head and hair region of Lena image is significantly visible in the SS-DCI$^3$ measurements compared to other methods. \textcolor{black}{From Fig. \ref{fig:learn_meas}. The order of increasing revealing information and increasing sampling efficiency is CSNet, PCS-DCI$^3$, W-DCI$^3$, and SS-DCI$^3$. This is the cost of improving performance by the learnable decomposition in SS-DCI$^3$.} 

\subsubsection{Linear vs. Non-linear in Multi-Scale Sampling}
The impact of linearity and bias term at the sampling stage was evaluated in Table \ref{tab:PerLinear}. The convolution layer was either (i) linear; (ii) linear with bias term; (iii) non-linear with ReLU and without bias term. While non-linearity often helps in autoencoder, it significantly reduces the reconstruction quality in all phases. As captured measurement can be negative to positive, using ReLU will set all negative measurements to zeros thus reduces the amount of information at the sampling stage. In contrast, adding bias increases parameters to the network thus slightly improved reconstruction quality. However, the bias term only showed improvement for simple networks at Phase 1 and Phase 2. With a complex reconstruction network like Phase 3, adding bias term in the sampling stage degraded the PSNR performance. Therefore, we conclude that the linearity is important at the sampling stage to preserve the input signal information. Therefore, we would suggest to use first few layers in a deep convolution network without the non-linear activation. This observation agreed with the linear bottlenecks model in MobileNetv2 \cite{MobileV2}.

 \begin{table}[]
 	\renewcommand{\arraystretch}{1.3}
 	\setlength{\tabcolsep}{2.5pt}	
 	\caption{Reconstruction performance of W-DCI in PSNR(dB) with Linear/ Linear \& Bias/ and Nonlinear sampling }
 	\label{tab:PerLinear}
 	\centering
 	\begin{tabular}{llc|c|c}
 		\hline
 		& rate & W-DCI$^1$            & W-DCI$^2$             & W-DCI$^3$             \\ \hline   
 		\multirow{2}{*}{Set5}  & 0.1  & 30.66/ 30.82/ 24.57 & 32.44/ 32.57/ 30.71 & 33.39/ 33.34/ 32.47 \\ \cline{3-5}
 		& 0.2  & 34.06/ 34.08/ 25.56 & 35.82/ 35.97/ 32.18 & 36.56/ 36.61/ 35.75 \\ \hline 
 		\multirow{2}{*}{Set14} & 0.1  & 27.81/ 29.90/ 23.27 & 29.10/ 29.16/ 27.81 & 29.67/ 29.56/ 28.91 \\ \cline{3-5}
 		& 0.2  & 30.69/ 30.69/ 23.81 & 32.05/ 32.19/ 28.79 & 32.51/ 32.45/ 31.02 \\ \hline  
 		
 	\end{tabular}
 \end{table}

\begin{table*}[]
	\renewcommand{\arraystretch}{1.3}
	\setlength{\tabcolsep}{2.2pt}	
	\caption{Number of parameters and variables of various networks, in Million.}
	\centering
	\begin{tabular}{lc|cc|cccc|cccc|cccc}
		\hline 
		&\multirow{2}{*}{rate} & \multirow{2}{*}{\scriptsize CSNet} & \multirow{2}{*}{\scriptsize SCSNet} & \multicolumn{4}{c|}{W-DCI$^i$}  &\multicolumn{4}{c|}{P-DCI$^i$} & \multicolumn{4}{c}{SS-DCI$^i$} \\ \cline{5-16}
		& &            &   &0     & 1      & 2      & 3      &0 & 1      & 2      & 3      &0 & 1       & 2      & 3       \\ \hline 
		\multirow{3}{0.7in}{parameters  (weights)} &0.1  & 0.321  & 0.340  &0.104       & 0.209   & 0.321    & 16.492  &0.039  & 0.143   & 0.255   & 16.426  & 0.027  & 0.033    & 0.145   & 16.316   \\
		&0.2  & 0.532  & 0.551  &0.210       & 0.420  & 0.532   & 16.703 &0.076  & 0.286   & 0.398   & 16.568  & 0.052    & 0.065    & 0.177   & 16.348   \\
		&0.3  & 0.741  & 0.760  &0.314       & 0.629   & 0.741   & 16.912  &0.114  & 0.429   & 0.541   & 16.711  & 0.079    & 0.099    & 0.211   & 16.381   \\ \hline 
		\multirow{3}{0.7in}{variables \\ (intermediate features)} &  0.1&135.03   & 371.48  &0.288   &0.813   & 135.29 & 544.24  & 1.101 & 1.625   &136.11   & 545.05  & 2.134  &2.648 & 137.13   & 546.07   \\
		&0.2& 135.06  & 372.03   & 0.315     &0.839  & 135.32   & 544.26  & 1.154  & 1.678   & 136.16  & 545.10  & 2.149  &2.674   &137.15   & 546.10    \\
		&0.3& 135.08  & 372.59  & 0.341     &0.865  & 135.35   & 544.29  & 1.206  & 1.730   & 136.21  & 545.15  &2.176  &2.700   &137.18   &  546.12    \\ \hline 
		\multicolumn{16}{r}{\scriptsize \textit{$i=0$ means the complexity of the sampling stage (without reconstruction)}}
	\end{tabular}
    \label{tab:params_vars}
\end{table*}

\begin{figure*}[!t]
	\centering
	\includegraphics[scale=0.33]{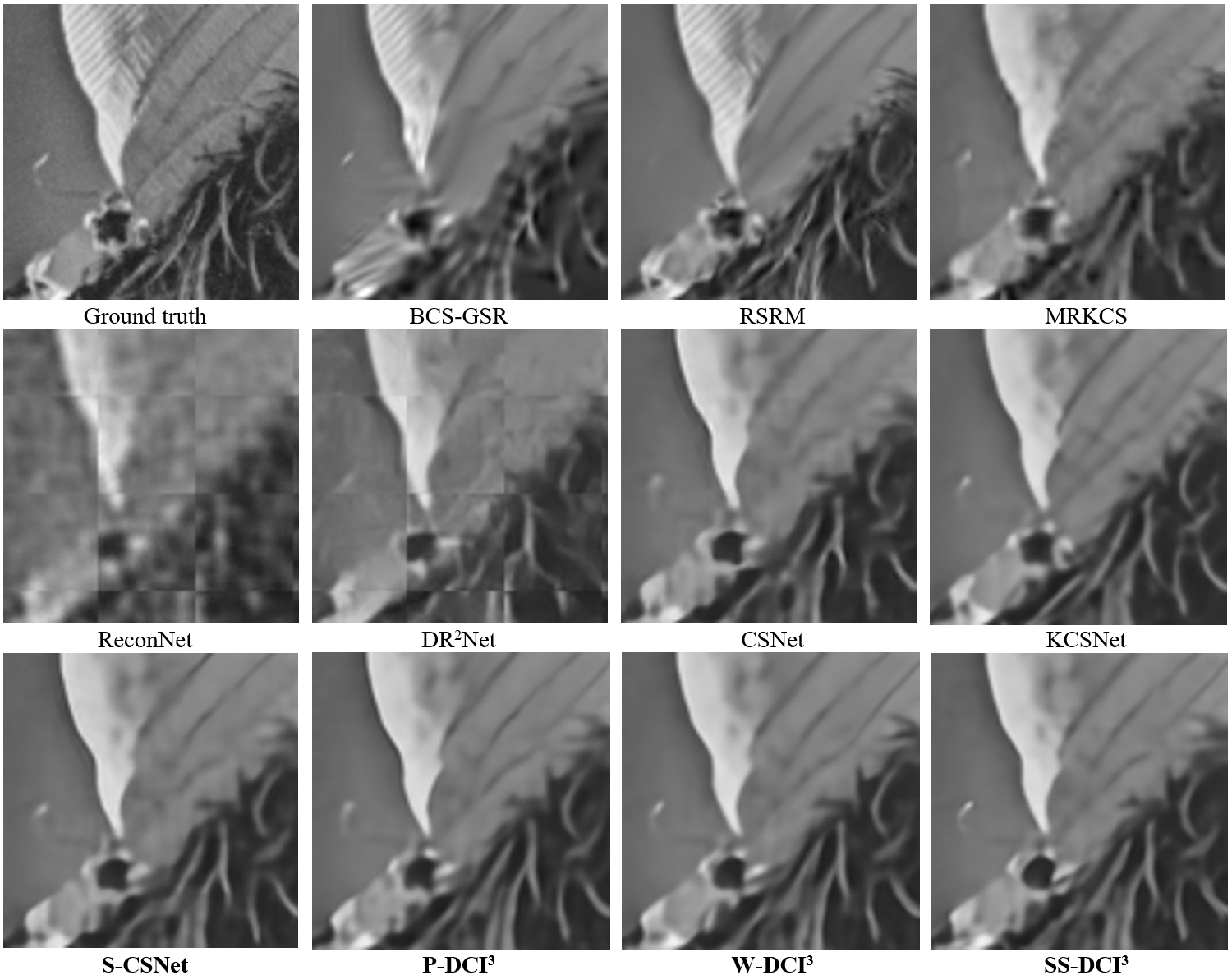}
	\caption{Cropped visual reconstruction of Lena image at subrate 0.1 of various sampling schemes. The vertical axis indicates the ranges of measurement.  }
	\label{fig:lena}
\end{figure*}

\subsection{Comparison with State-of-the-Art Sampling Schemes}
\subsubsection{\textcolor{black}{Reconstruction Quality}}
For Set5 and Set14 in Table \ref{tab:Per01}, all multi-scale networks outperform the single scale sampling scheme. While previous MS-DCI works (MS-CSNet and S-CSNet) produces similar performance, they showed an average improvement of 0.50$\sim$0.62 dB over the single scale DCI (CSNet and DBCS) and 1.98$\sim$2.79 dB over the conventional CS (MH and GSR). Our proposed MS-DCIs showed steady improvement over each training phase. At Phase 3, SS-DCI$^3$, W-DCI$^3$, P-DCI$^3$ gained more than 3.33dB, 3.09dB, 2.97dB over CSNet on Set5 at subrate 0.1.  Moreover, our SS-DCI$^3$ outperformed MS-CSNet and S-CSNet with 0.47$\sim$0.86dB and 0.39$\sim$0.47dB gain on Set5 and Set14, respectively.  \textcolor{black}{Both W-DCI$^3$ and P-DCI$3$ improved 0.45$\sim$0.62 dB and 0.33$\sim$0.50 dB over S-CSNet at subrate 0.1. W-DCI$^3$ still offered higher performance than MS-CSNet and S-CSNet at high subrate. P-DCI$3$ resulted in better PSNR at subrate 0.2 and similar or less at subrate 0.3.} In general, the smaller subrate is (or the more ill-posed the problem is), the higher improvement in reconstruction is achieved by our MS-DCIs. 
\textcolor{black}{Overall, the order of reducing performance are SS-DCI$^3$, W-DCI$^3$, P-DCI$^3$, MS-CSNet, S-CSNet, SS-DCI$^2$, W-DCI$^2$, P-DCI$^2$. }

\begin{figure*}[!t]
	\centering
	\includegraphics[scale=0.33]{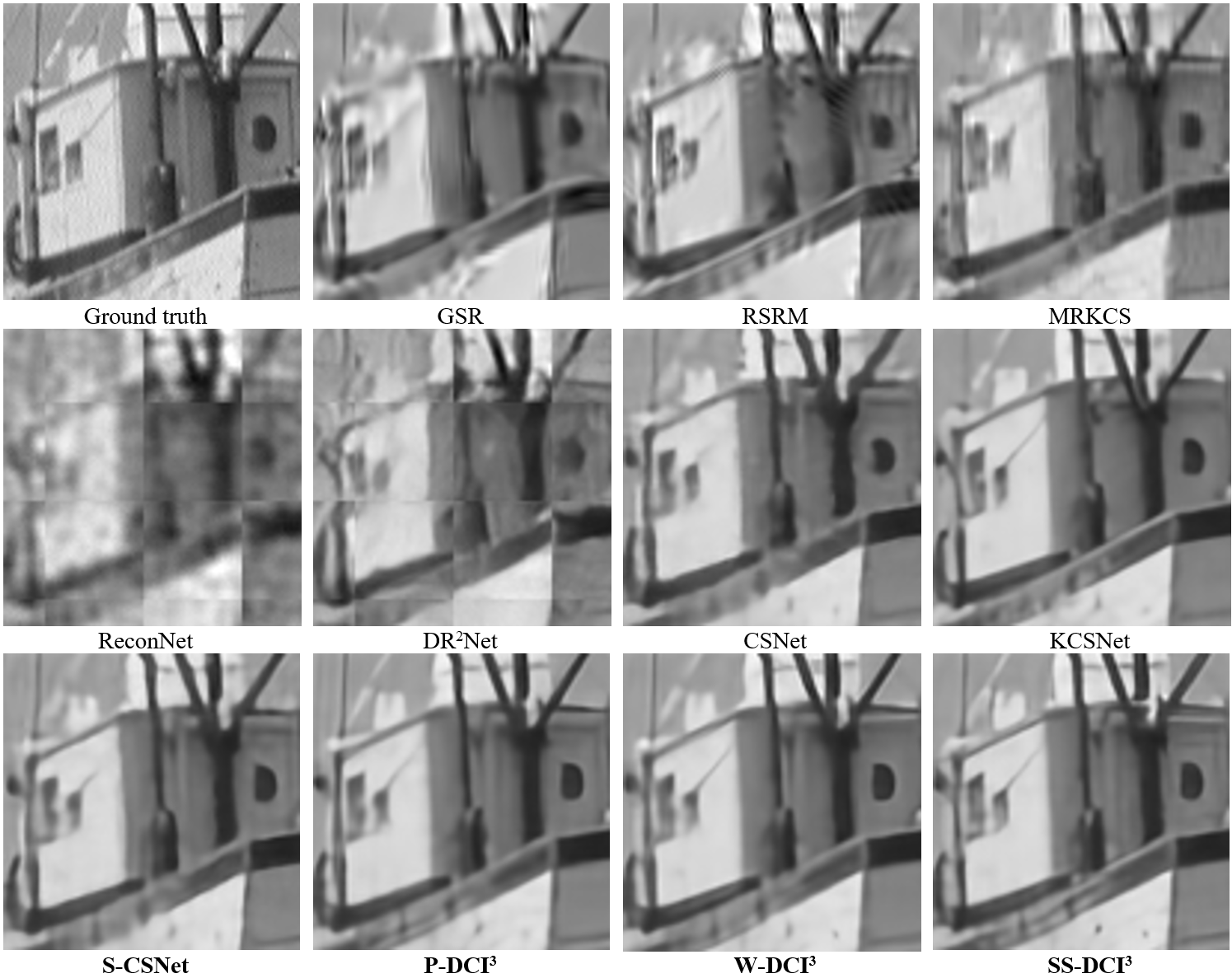}
	\caption{Cropped visual reconstruction of Boats image at subrate 0.1  with various sampling schemes.   }
	\label{fig:boats}
\end{figure*}

\begin{figure}[!t]
	\centering
	\includegraphics[scale=0.63]{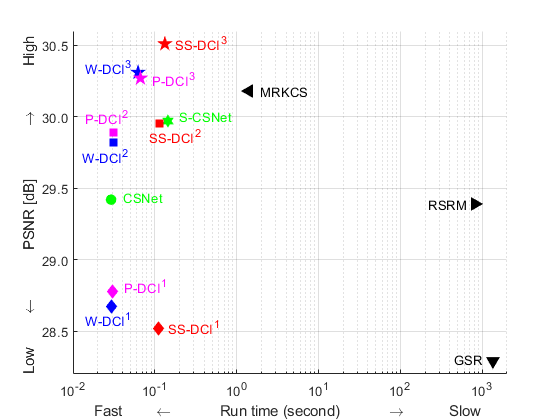}
	\caption{Running time (seconds) vs. reconstruction quality in average (PSNR[dB]) for six $512\times 512$ test images at subrate 0.1. SS-DCI$^3$ shows the best performance with limited additional complexity.}
    \label{fig:runingtime}
\end{figure}

For $512\times512$ images in Table \ref{tab:PerSOTA}, without learning the sampling matrices, ReconNet and DR$^2$Net resulted in 3.88dB less than the conventional CS with RSRM and other jointly learning reconstruction and sampling networks. Single scale DCI, CSNet, showed 0.4$\sim$1.5 dB gain over GSR but less than the conventional single scale RSRM (i.e. around 0.66$\sim$0.84 dB) and multi-scale CS with MRKCS (-0.03$\sim$0.89 dB). Our multi-scale networks (P-DCI$^3$, W-DCI$^3$, and SS-DCI$^3$) outperformed both single-scale CSNet and KCSNet, and multi-scale S-CSNet. Thanks to the jointly learn of decomposition, sampling, and reconstruction, SS-DCS$^3$ demonstrated the best reconstruction with up to 6.83 dB gain over CSNet and S-CSNet for the highly compressible image (i.e. Cameraman) at subrate 0.3. For complex images like Lena, Mandrill, SS-DCI$^3$ still showed 0.36$\sim$0.93 dB gain over CSNet. Compared to SS-DCI$^3$, W-DCI$^3$ and P-DCS$^3$ shown comparable and less reconstruction quality at the high subrate. 

Visual quality is presented in Fig. \ref{fig:lena} and \ref{fig:boats}. Conventional CS (GSR and RSRM) preserved the high-frequency detail regions well (i.e., Lena’s hat) but created the fake edge artifacts in complex regions (i.e. Boats and Lena’s hair). It is because the low-rank assumption does not hold for the complex texture regions. 
In contrast, losing high frequency was observed in the conventional multi-scale CS (i.e, MRKCS) and all DCI methods, especially for strong local structure regions like Lena’s hat. The effective of our multi-scale sampling is demonstrated by the best visual quality of SS-DCI$^3$ following by W-DCI$^3$ and P-DCI$^3$. Other multi-scale method S-CSNet, KCSNet, and MRKCS surfer aliasing artifact. 

\subsubsection{Complexity}
\textcolor{black}{The complexity of DCI is expressed by the number of learning parameters (i.e. learning capability) and the number of variables (i.e. size of the intermediate features) in Table \ref{tab:params_vars}. Generally, sampling requires much less computation than reconstruction, especially at Phase 3. Complexity and performance of our MS-DCIs are  increased after each reconstruction phase ($i=1,2,3$). Meanwhile, the fixed decomposition of W-DCI needed more parameters but less variables than P-DCI and SS-DCI.} 

\textcolor{black}{At Phase 2, our MS-DCIs required a similar number of parameters (number of weights) and variables (number of intermediate features) as the single-scale CSNet but offered 0.15$\sim$0.30dB improvement in PSNR. S-CSNet improved reconstruction quality by slightly increasing parameters and double variables. At phase 3, we added $30 \times $ parameters and $3.5\times$ variables over Phase 2. Also, compared to S-CSNet, our MS-DCIs were 30$\times$ more parameters but only $1.5\times $ variables. One might conclude that the PSNR improvement of our proposed MS-DCIs has solely come from the multi-scale reconstruction than the sampling method. However, at the same reconstruction complexity, only a slight difference in sampling complexity, SS-DCI$^3$ still gained 0.20dB over W-DCI$^3$ and P-DCI$^3$. Therefore, we concluded that sampling architecture has a significant impact on the final reconstruction performance. Also, a reduction in parameters comes with the cost of increasing variables.} 

\textcolor{black}{ For the complexity, we showed the running times (in seconds) versus the average reconstruction quality (in PSNR [dB]) for six images in Fig. \ref{fig:runingtime} wit a PC system running Windows 10 Home Edition, Matlab 2019a, 32Gb Ram, GPU Nvidia 2080ti, and MatConvNet1.0 beta 25. DCI schemes were tested with 1000 average for each image and the conventional CS schemes (MRKCS, RSRM, and GSR) are averaged 10 times. Conventional single CS (RSRM and GSR) showed poor quality and slow reconstruction. Conventional multi-scale CS (MRKCS) offered higher quality and faster reconstruction but still significantly slower than DCIs. SS-DCI is always slower than W-DCI and P-DC but better in performance at Phase 2 and 3. It clearly showed that W-DCI$^2$ and P-DCI$^2$ were better than CSNet with the same running time. S-CSNet had similar PSNR at the same running time as SS-DCI$^2$ and slower than P-DCI$^2$. All MS-DCI at Phase 3 were better than S-CSNet in performance and running time.}

\subsection{\textcolor{black}{Discussions and Future Work}}


\textcolor{black}{ One of the most obvious application of our multi-scale framework is \textit{learned image/video compression} \cite{Nakasihi18} which compressing images at single-scale. Since it is known that multi-scale processing benefits compressive sensing and image compression in JPEG2000 \cite{David12}, jointly learn multi-scale decomposition, multi-scale image compression, will further improve the compression ratio. The second research topic that could utilize the multi-scale sampling concept is \textit{coded imaging}. It is possible to model the multi-scale sampling via multi-level of exposures. For instance, some pixels are exposed shorter to capture fast moving object (high frequency) while others are exposed longer to capture stationary object (low frequency). The third application \textit{MRI sampling and reconstruction} \cite{Dai19}. Since the MRI image is captured in the Fourier decomposed domain, it is possible to mimic the Fourier transform and radial sampling as a layer to enable end-to-end multi-scale sampling.}

\textcolor{black}{Our MS-DCI offers an alternative explanation for the multi-level wavelet convolution (MWCNN) \cite{Liu19}. In MWCNN, the authors decomposed features independently by wavelet before performing convolution across multiple channels. MWCNN was interpreted as an efficient pooling method to avoid the gridding artifact of the conventional pooling. However, as a convolution layer also can be interpreted as a CS sampling scheme with linear projection (see Fig. \ref{fig:dl_vs_bcs}). In our view, we can consider the conventional convolution (CNN) and MWCNN as single-scale and multi-scale sampling with wavelet, respectively. As we demonstrate the advantage of multi-scale over single scale sampling, MWCNN can capture multi-scale features more efficiently than conventional CNN, thus archives higher reconstruction performance. Additionally, our results indicated that wavelet is not the best decomposition in multi-scale sampling. With that assumption, scale-space CNN would show better performance than MWCNN.}  

\section{Conclusion}
This work proposed a novel multi-scale deep learning-based compressive imaging network to improve the sampling efficiency and reconstruction quality. The proposed framework not only learns to decompose and sample images at multi-scales but also reconstruct images at multi-scale. We demonstrated the importance of sampling networks in improving the final reconstruction performance with merely additional complexity. We proposed a three-phase training scheme to further improves training efficiency and reconstruction quality. The characteristic of learned decomposition, learning multi-scale sensing were investigated including Pyramid, Wavelet, and Scale-space decomposition. After the training process, our multi-scale sampling matrices are fixed and can be applied for sequential imaging systems.  


%

\appendices


\section*{Acknowledgment}

This work is supported in part by the National Research Foundation of Korea (NRF) grant 2017R1A2B2006518 funded by the Ministry of Science and ITC.

\ifCLASSOPTIONcaptionsoff
  \newpage
\fi

\end{document}